\documentclass[sort&compress,twocolumn,3p]{elsarticle}

\usepackage{mathtools}
\usepackage{gensymb}
\usepackage{graphicx}
\usepackage{xfrac}

\begin{document}

\title{Relative Cooling Power Enhancement by Tuning Magneto-structural Stability in Ni-Mn-In Heusler Alloys}

\author[1]{Jing-Han~Chen}
\ead{jhchen10@lsu.edu}
\author[2]{Nickolaus~M.~Bruno}
\ead{nickolaus.bruno@sdsmt.edu}
\author[1]{Zhenhua~Ning}
\author[3]{William~A.~Shelton}
\author[4,5]{Ibrahim~Karaman}
\author[6]{Yujin~Huang}
\author[6]{Jianguo~Li}
\author[7,4]{Joseph~H.~Ross,~Jr.}

\address[1]{Department of Physics and Astronomy, Louisiana State University, Baton Rouge, Louisiana 70803, USA}
\address[2]{Department of Mechanical Engineering, South Dakota School of Mines and Technology,	Rapid City, South Dakota 57701, USA}
\address[3]{Department of Chemical Engineering, Louisiana State University, Baton Rouge, Louisiana 70803, USA}
\address[4]{Department of Materials Science and Engineering, Texas A\&M University, College Station, Texas 77843, USA}
\address[5]{Department of Mechanical Engineering, Texas A\&M University, College Station, Texas 77843, USA}
\address[6]{School of Materials Science and Engineering, Shanghai Jiaotong University, Shanghai, 200240, China}
\address[7]{Department of Physics and Astronomy, Texas A\&M University, College Station, Texas 77843, USA}

\date{\today}

\begin{abstract}
Off-stoichiometric Ni$_2$MnIn Heusler alloys have drawn recent attention due to their large magnetocaloric entropy change associated with the first-order magneto-structural transition.
Here we present crystal structural, calorimetric and magnetic studies of three compositions.
Temperature-dependent X-ray diffraction shows clear structural transition from a 6M modulated monoclinic to a L2$_1$ cubic.
A significant enhancement of relative cooling power (RCP) was achieved by tuning the magnetic and structural stability through minor compositional changes, 
with the measured results quantitatively close to the prediction as a function of the ratio between the martensitic transition ($T_m$) temperature and austenite Curie temperature ($T_C$)
		although the maximal magnetic induced entropy change ($\Delta S_{max}$) reduction is observed in the same time.
The results provided an evaluation guideline of RCPs as well as magnetic-induced entropy change in designing practical active materials.
\end{abstract}


\maketitle

\section{Introduction}
Materials showing the magnetocaloric effect (MCE) have been a source of growing interest because of their 
potential for an environmentally friendly and energy efficient refrigeration technology.
Recently, many of the off-stoichiometric Heusler alloys based on
Ni-Mn-Z (Z=In, Sb, Sn) have been reported to show a large MCE near room temperature\cite{tishin2003magnetocaloric,doi:10.1063/1.4874935,doi:10.1063/1.1808879,krenke2005inverse,doi:10.1063/1.2187414,:/content/aip/journal/jap/101/5/10.1063/1.2710779,Liu2012514,Bennett201234}.
In particular, Ni-Mn-In type alloys have been of interest for their large MCE͒
in the vicinity of first-order magneto-structural phase transitions, and hence have potential application as a working material in magnetic regenerators\cite{PhysRevB.75.104414,:/content/aip/journal/apl/89/18/10.1063/1.2385147}.
Composition maps in the Ni-Mn-In system have been recently developed lending the ability to design materials\cite{:/content/aip/journal/apl/101/22/10.1063/1.4768235,Chen2016176,:/content/aip/journal/jap/116/20/10.1063/1.4902527}.

\begin{figure}
\includegraphics[width=\columnwidth]{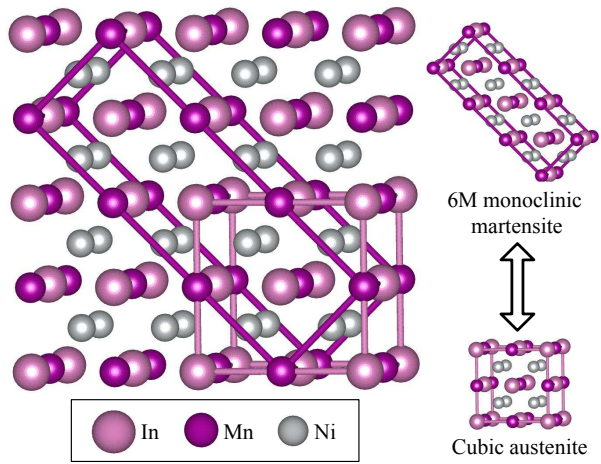}
\caption{2$\times$2 set of L2$_1$ stoichiometric Ni$_2$MnIn cubic unit cells is shown to illustrate the crystal structure of the monoclinic martensite phase as well as the cubic austenite phase.
Illustration shows unit cells for both cases, with 6M structure coming from the monoclinic distortion of the cubic austenite supercell along (1 1 0) direction.
In this report, off-stoichiometric Ni$_2$MnIn samples were prepared so that the concentration ratio of Mn:In is not 1:1.
}
\label{intro}
\end{figure}

Since the Ni-Mn-In Heusler alloys undergo a first-order magneto-structural phase transition, which is intimately related to the change of the long-range crystal symmetry,
the crystal structures of the austenite and martensite phases have been widely investigated experimentally\cite{Yan2015375,PhysRevB.73.174413,HERNANDO2009763,doi:10.1063/1.2827179,doi:10.1063/1.3574088,doi:10.1063/1.1808879,PhysRevB.75.104414,doi:10.1063/1.2187414,doi:10.1063/1.2913162,LIU20094911,KenichiAbematsu2014M2013372}.
It has been shown that the crystal structures of the martensite phase in Ni-Mn-In Heusler alloys are very sensitive to chemical compositions as well as material processing
while the crystal structure of austenite phase remains cubic over a wide range of composition.
In this report, temperature-dependent X-ray diffraction spectra are reported.
The crystal structures of both martensite and austenite are identified and connected to their magnetocaloric properties.
The high-temperature austenite phase is cubic while the low-temperature martensite phase is 6M modulated monoclinic.
The crystal structure of the modulated 6M monoclinic martensite phase can be described as the monoclinic distortion of the $3\times1$ supercell of the austenite phase along (1 1 0) direction.
The relation of the crystal structures between the martensite and austenite phase is illustrated in Fig.~\ref{intro} in case of the stoichiometric Ni$_2$MnIn.

One of the common physical quantities used to compare MCE materials is the relative cooling power (RCP),
	which represents the amount of transferred heat between hot and cold reservoirs in a magnetic refrigeration cycle\cite{Pecharsky199944,franco2012magnetocaloric}.
We previously presented an analytic model, which was in good quantitative agreement with the experimental results,
and predicted RCP enhancement by reducing the ratio between the martensitic transition and austenite Curie temperatures\cite{:/content/aip/journal/jap/116/20/10.1063/1.4902527}.
In this paper, we prepared off-stoichiometric Ni-Mn-In based Heusler alloys with three different compositions.
As the concentration of Mn increased, the strength of atomic magnetic interaction is reduced so that the magnetism of the austenite phase was modified from ferromagnetic to paramagnetic.
At the same time, the martensitic structural transition temperature increased.
The experimental RCPs for these materials are in a good quantitative agreement with the proposed model.
By varying the ratio between martensitic transition and Curie temperature of austenite phase,
   we achieved a significant RCP enhancement while the maximal magnetic induced entropy change was reduced.

\section{Experimental Procedures}
\subsection{Sample Preparation}
Bulk polycrystalline Ni-Mn-In alloys were prepared using arc melting in a protective high purity argon atmosphere from 99.9\% pure constituents.
The samples were homogenized at 900 $\celsius$ for 24 hours under low pressure argon atmosphere and then quenched in room-temperature water.
After heat treatment, the ingot was then cut into pieces for different measurements.
These heat treating conditions were found to result in single phase martensite or austenite across sharp first-order transitions as shown below.

The nominal compositions of these three samples were Ni$_{50}$Mn$_{36}$In$_{14}$ (sample A), Ni$_{50}$Mn$_{35.5}$In$_{14.5}$ (sample B) and Ni$_{50}$Mn$_{35}$In$_{15}$ (sample C).
The final compositions of these three samples were measured by four wavelength-dispersive X-ray spectrometers (WDS) in a Cameca SX50 electron microprobe, and are Ni$_{49.54}$Mn$_{36.12}$In$_{14.34}$ (sample A), Ni$_{49.88}$Mn$_{35.70}$In$_{14.42}$ (sample B) and Ni$_{49.53}$Mn$_{35.22}$In$_{15.22}$ (sample C) as tabulated in Table~\ref{comptable}.
In order to confirm the compositional homogeneity of the annealed samples, we performed the measurement in 5-10 different grains and the composition variation between the grains in each sample was less than 0.5 atomic \%
while the errors due to the elemental detection limit of the WDS measurements within the microprobe were 0.047 atomic \% for Ni, 0.024 atomic \% for Mn and 0.007 atomic \% for In.

\begin{table*}
\centering
\begin{tabular}{l|ccccccccc}
\hline\hline
Label&WDS composition&$J$&$T_{m}(K)$&$T_C(K)$&$\sfrac{T_m}{T_C}$ & $\Delta$S$_{max}$(mJ/g-K)&$\sfrac{\Delta V}{V_{100K}}(\%)$\\
\hline
A	&Ni$_{49.54}$Mn$_{36.12}$In$_{14.34}$	&2.00	&347	&292	&1.19 &	50.1 & 1.97 \\
B	&Ni$_{49.88}$Mn$_{35.70}$In$_{14.42}$	&1.99	&333	&310	&1.07 &	44.0 & 1.87 \\
C	&Ni$_{49.53}$Mn$_{35.22}$In$_{15.22}$	&2.00	&299	&323	&0.93 &	30.4 & 1.74 \\
\hline\hline
\end{tabular}
\caption{The corresponding experimental results of our three samples are shown.
		$J$ and $T_C$ are estimated from the high-temperature Curie-Weiss fits, and
		$T_{m}$ (martensitic transition temperature) is defined as the maximum position of the specific heat measured while heating.
The volume changes, $\Delta V=V_{400K}-V_{100K}$, are determined from the X-ray diffraction.}
\label{comptable}
\end{table*}

\subsection{X-ray diffraction}
\begin{figure}
\includegraphics[width=\columnwidth]{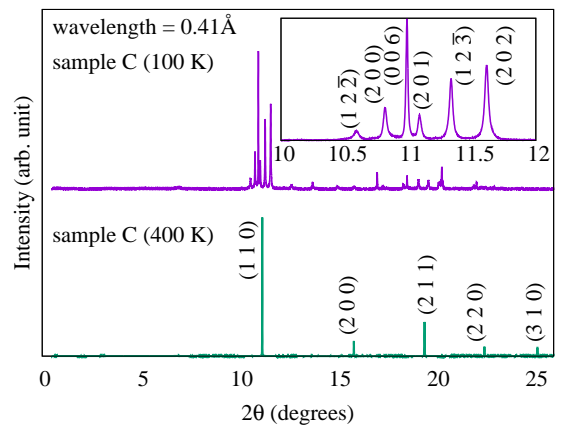}
\caption{X-ray diffraction at 100 K and 400 Kelvin results for the sample C are shown.
The (1 1 0) intensities for all three samples are normalized to be equal to each other for comparison.
The domination of cubic $B2$ structure in austenite phases is confirmed.
Low temperature (martensite phase) X-ray diffraction results agree well with the reported 6M modulated structural parameters (space group: $P2/m$)\cite{Yan2015375}.}
\label{xrd}
\end{figure}

Temperature-dependent high resolution powder X-ray diffraction (XRD) data were collected at the X-ray Science Division beamlines at the Advanced Photon Source, Argonne National Laboratory by using a X-ray wavelength of $~0.41$\AA.
After grinding to powders, these samples were heat treated using the same methods as the bulk in order to relax the lattice strain from the grinding process.
The temperature-dependent XRD patterns for all three samples show that they have the same martensite crystal structure and austenite crystal structure.
Representative results for Sample C in the temperature region of the martensite and austenite phases are shown in Fig.~\ref{xrd}.
The signature (1 1 1) reflection at $2\theta=6.8$ degrees of the Heusler-type $L2_1$ order in austenite is absent in Fig.~\ref{xrd},
	which confirms the statistical disorder occupancy between Mn and In atoms.
Hence we conclude the domination of cubic $B2$ structure and this is consistent with previous studies\cite{Recarte20121937,Bruno2017}.

Rietveld refinements were performed using GSAS software\cite{larson94,toby01} and the lattice constants from the refinement are tabulated in Table~\ref{xrdtable}.
The results in the temperature region of the martensite phase (100 K) show the 6M modulated structure (illustrated in Fig.~\ref{intro}, space group: $P2/m$),
	as previously reported\cite{Yan2015375,KenichiAbematsu2014M2013372,ADFM:ADFM200801322,doi:10.1063/1.3043456} while 10M and 14M modulated structures for similar compositions have also been reported\cite{PhysRevB.73.174413,doi:10.1063/1.2827179,0953-8984-21-23-233201}.
Our results indicate that the 6M martensite structure persists for samples tuned across this compositional range through which the austenite changes from ferromagnetic to paramagnetic just above the structural transition,
	even though the lattice contribution to the transformation entropy has been shown to undergo a large change across this range\cite{Chen2016176,PhysRevB.92.140406}.

As the temperature rises, these crystal structures transform from monoclinic to cubic.
We observe similar cubic lattice constants for the austenite phase (400 K) of three samples as shown in Table~\ref{xrdtable}.
The resulting volume changes across the martensitic transition from 100 K to 400 K are also shown in Table~\ref{comptable}.
These changes follow the ordering A $>$ B $>$ C, which is in line with the maximal magnetic induced entropy estimated from the specific heat shown later.

\begin{table}
\centering
\footnotesize
\begin{tabular}{l|p{8mm}p{8mm}p{9mm}p{6mm}p{8mm}}
\hline\hline
samples	&	$a$(\AA)	&	$b$(\AA) &	$c$(\AA)	& $\beta$(\degree) & vol.(\AA$^3$)\\
\hline
A,100 K	&4.407(7)	&5.571(1)	&13.000(3)	&93.5(0)	&	13.276(8)\\
B,100 K	&4.408(7)	&5.574(5)	&13.006(6)	&93.5(3)	&	13.294(1)\\
C,100 K	&4.401(7)	&5.600(7)	&13.006(6)	&93.0(7)	&	13.341(4)\\
\hline
A,400 K	&3.002(8)	&3.002(8)	&3.002(8) 	&90.0		&	13.538(2)\\
B,400 K	&3.003(2)	&3.003(2)	&3.003(2) 	&90.0		&	13.543(3)\\
C,400 K	&3.005(5)	&3.005(5)	&3.005(5) 	&90.0		&	13.573(9)\\
\hline\hline
\end{tabular}
\caption{The lattice parameters at 100 K and 400 K from the Rietveld refinement. volume in this table is defined in the unit of \AA$^3$ per atom.
		The austenite phases (i.e. 400 K) are crystallized in disorder cubic $B2$ structure (space group: $Pm\bar{3}m$) while the martensite phase (i.e. 100 K) are crystallized in 6M modulated monoclinic structure (space group: $P2/m$).}
\label{xrdtable}
\end{table}

\subsection{Magnetic Experiments and Calculation}
\begin{figure}
\includegraphics[width=\columnwidth]{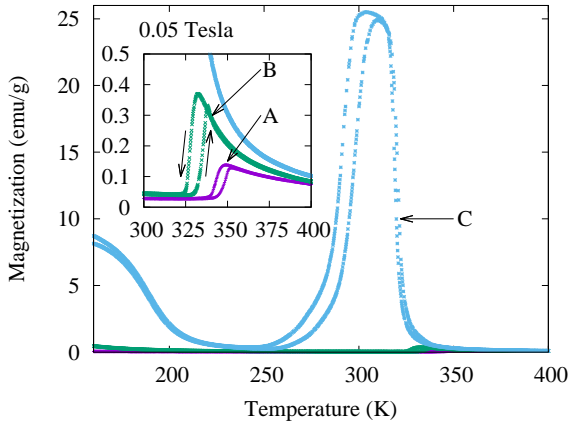}
\caption{Temperature dependence of the magnetization for the sample A, B and C at 0.05 T is shown.
All data include results for both heating and cooling processes, as shown by arrows for	sample B in the inset figure.}
\label{mt}
\end{figure}

Iso-field magnetization measurements were carried out using a Magnetic Property Measurement System (MPMS, Quantum Design) on a sample from the center of the homogenized sample plate.
Figure~\ref{mt} shows the temperature-dependent magnetization in 0.05 Tesla.
The transition curves, as shown, correspond to the martensitic transformation. 
The forward (austenite to martensite) transformation is observed on cooling the samples and the reverse transformation (martensite to austenite) is observed on sample heating. 
The results indicate a transition from a ferromagnetic/paramagnetic austenite to a low-magnetization martensite upon cooling.
However, as we previously demonstrated in Ref.~\cite{Chen2016176}, the martensite phases have a relatively high degree of spin ordering (and hence reduced spin entropy) in a configuration characterized as antiferromagnetic.

The Curie-Weiss law,
\begin{equation}
M=n\frac{N_A}{3k_B}\mu_{eff}^2\frac{H}{T-T_C},
\end{equation}
where $\mu_{eff}=g\mu_B\sqrt{J(J+1)}$, $T_C$ is the Curie temperature and $g=2$,
was used to fit the high temperature paramagnetic magnetization curves.
This fitted spin value was obtained by assuming that the density of magnetic moments ($n$) is identical to that of the manganese ions.
The fitted results for $T_C$ and $J$ are listed in Table~\ref{comptable}.
All of these results are in agreement with a local magnetic moment of $4\mu_B$ in line with reported studies\cite{Li201335,0953-8984-21-23-233201,PhysRevB.86.214205,:/content/aip/journal/apl/97/24/10.1063/1.3525168} 
and are consistent with local magnetic moments in the sample attributed to manganese with negligible moment on other ions.
It is interesting that for Ni-Co-Mn-Sn materials\cite{doi:10.1063/1.4960353} the magnetism exhibits behavior much closer to itinerant magnetism, however in these materials the identical per-Mn moments point to magnetism that is very well represented as composed of local moments on Mn ions.
The order of the ratio between martensitic transition and austenite Curie temperature ($\sfrac{T_{m}}{T_C}$) is A $>$ B $>$ C as shown in Table~\ref{comptable}.

It is worthy noting that the Curie temperature $T_C$ decreases as the increasing of the compositional concentration of Mn, showed in Table~\ref{comptable}.
To understand this behavior qualitatively, we calculated the $T_C$ of two binary compounds MnNi and MnNi$_3$ by mapping energy differences onto the general Heisenberg Spin Hamiltonian within density functional theory\cite{Hohenberg1964,Kohn1965} using the projector augmented wave method\cite{Kresse1999} implemented in the Vienna ab initio simulation package\cite{Kresse1993,Kresse1996}.
The generalized-gradient approximation to the exchange-correlation functional proposed by Perdew, Burke and Ernzerhof\cite{Perdew1996} was used.
By adapting the L2$_1$ austenite cubic crystal structure, we obtained that the $T_C$ of MnNi was less than that of MnNi$_3$ which explains why $T_C$ was decreased as the concentration of Mn atoms is increased in our materials.
This gives us some insight that the magnetic exchange coupling is weakened as Mn concentration increases.

\subsection{Entropy from Specific Heat Measurements}
\begin{figure}
\includegraphics[width=\columnwidth]{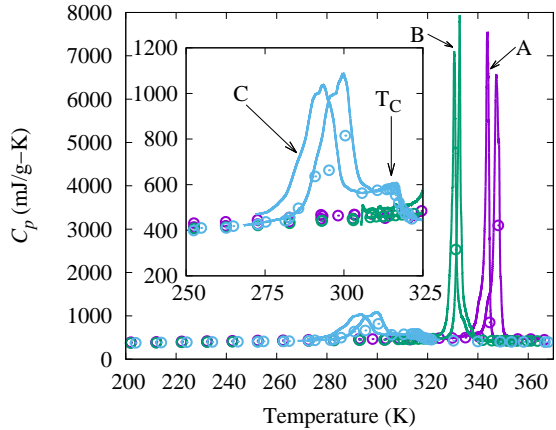}
\caption{Solid curves: specific heat obtained from heating and cooling measurements analyzed as described in text.
		Open symbols: results from the conventional 2-$\tau$ analysis, with small heat pulses, showing good agreement outside of first-order transition regime.
The inset figure shows the magnification of specific heat results from sample C and its Curie temperature ($T_C$) is labelled.
}
\label{Cp}
\end{figure}

The absolute total entropy of the material can be estimated directly from the integration of the specific heat data
\begin{equation}
S(T, H=0)=\int_0^T\frac{C_p(T^\prime,H=0)}{T^\prime}dT^\prime.
\label{S}
\end{equation}
Specific-heat measurements were performed using a Physical Property Measurement System (PPMS, Quantum Design).
For temperatures away from the first-order martensitic transition, the conventional 2-$\tau$ method\cite{:/content/aip/journal/rsi/68/1/10.1063/1.1147722} was used.
For temperatures across the first-order martensitic transitions, the long heating and cooling pulses were used to guarantee the completion of the martensitic transition\cite{:/content/aip/journal/jap/116/20/10.1063/1.4902527,Chen2016176}.
The temperature-dependent specific heat results across the martensitic transition are shown in Fig.~\ref{Cp}.

\begin{figure}
\includegraphics[width=\columnwidth]{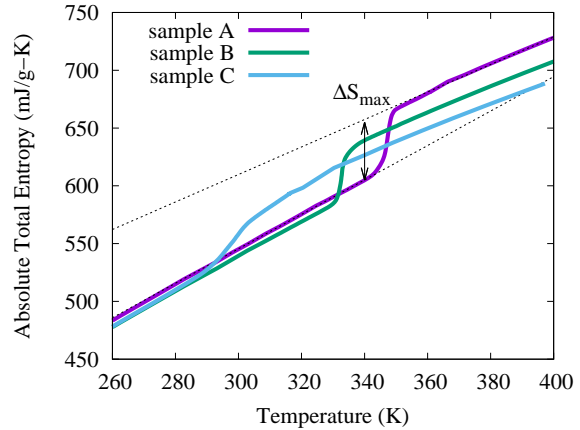}
\caption{The absolute entropy of sample A, B, C by Eq.~\ref{S} from the heating specific heat measurement are shown.
		The maximal magnetic-induced entropy $\Delta S_{max}$ of sample A is shown as a representative and the dash lines are from the linear fittings of the martensite phase entropy and austenite entropy to interpolate $\Delta S_{max}$.
		The magnitude order of $\Delta S_{max}$ across martensitic transition is A $>$ B $>$ C, which is consistent with the volume changes yielded from the X-ray diffraction above.}
\label{hysteresis}
\end{figure}

By utilizing Eq.~\ref{S} and the specific heat results shown in Fig.~\ref{Cp}, the absolute total entropy close to the martensitic transition were estimated as shown in Fig.~\ref{hysteresis}.
Therefore, the maximal magnetic induced entropy ($\Delta S_{max}$) was able to be determined from the entropy analysis by extending and interpolating the straight line shown in Fig.~\ref{hysteresis}.
The results show $\Delta S_{max}$ across the first-order martensitic transition to be A $>$ B $>$ C, which is consistent with the volume changes yielded from the X-ray diffraction data above.

Given that the lattice entropy can be approximated by the expression for the temperature region greater than the Debye temperature $\Theta$\cite{PhysRevB.71.054410}
\begin{equation}
S_L\approx3R\left(\ln T -\ln \Theta + \frac{4}{3}\right).
\label{S_L}
\end{equation}
For a first-order magneto-structural transition, this lattice entropy is connected with the change of the Debye temperature due to the first-order structural deformation\cite{PhysRevB.71.054410,doi:10.1063/1.2201879,doi:10.1063/1.3257381}
\begin{equation}
\Theta_A=\Theta_M\left(1-\gamma\frac{\Delta V}{V} \right),
\label{thetaA}
\end{equation}
where $\Theta_M$ is the Debye temperature of the martensite phase, $\Theta_A$ is the Debye temperature of the austenite phase and $\gamma$ is Gr\"{u}neisen parameter.
By utilizing Eq.~\ref{S_L} and Eq.~\ref{thetaA}, we can estimate the lattice entropy change due to the change of the Debye temperature.
Usually, $\gamma$ is between 1 and 3\cite{kittel2004introduction}.
By adopting the $\gamma=2.405$ from the Ni-Mn-Sb alloy by the first principle\cite{PUGACZOWAMICHALSKA2006251}, the resulting lattice entropy changes due to the change of Debye temperature are 4.63 mJ/g-K, 4.38 mJ/g-K and 4.04 mJ/g-K for sample A, B and C respectively.
Note that the calculated lattice entropy difference among the samples is much lower than the difference of $\Delta S_{max}$ in table~\ref{comptable}.
Therefore, the major $\Delta S_{max}$ difference among samples should be from the different magnetic order change across martensitic transition.

\section{Relative Cooling Power}
An important quantity for evaluating the performance of MCE materials is the amount of transferred heat
between cold and hot reservoirs in an ideal refrigeration cycle\cite{:/content/aip/journal/jap/90/9/10.1063/1.1405836,gschneidner1999recent}.
This is referred to as the relative cooling power (RCP).
The estimation of the RCPs come from the integration of the isothermal entropy change
\begin{equation}
\text{RCP}(H)=\int_{T_{cold}}^{T_{hot}}\Delta S(T, 0\to H)dT,
\label{rcp}
\end{equation}
where $T_{cold}$ and $T_{hot}$ are the temperatures of the two reservoirs.

Accurate determination for the isothermal entropy change can be obtained 
from either specific heat in different magnetic fields\cite{PhysRevB.77.214439}
or field-dependent magnetization in different temperatures by the relation\cite{PhysRevB.64.144406,Amaral20101552},
\begin{equation}
\begin{split}
&\Delta S(T, 0\to H)=\frac{\partial}{\partial T}\left(\int_0^H M(T,H^\prime)dH^\prime\right)_{T=const.}\\
&\cong\frac{1}{\Delta T}\left[\int_0^HM(T+\Delta T,H^\prime)dH^\prime-\int_0^HM(T,H^\prime)dH^\prime\right].
\end{split}
\label{mhentropyeq}
\end{equation}
These two methods have proven to be in good agreement recently for Ni-Mn-In Heusler alloys
given that the magnetic and calorimetric experiment were performed accurately\cite{0022-3727-45-25-255001,:/content/aip/journal/jap/116/20/10.1063/1.4902527}.

Furthermore, Eq.~\ref{rcp} can be rewritten\cite{Amaral20101552} by using Eq.~\ref{mhentropyeq}
\begin{equation}
\begin{split}
&\text{RCP}(H)=\int_{T_{cold}}^{T_{hot}}\Delta S(T, 0\to H)dT\\
&=\int_0^H M(T_{hot},H^\prime) dH^\prime-\int_0^H M(T_{cold},H^\prime) dH^\prime
\end{split}
\label{rcpmh}
\end{equation}
which indicates that RCP can be determined from isothermal magnetic measurements at only two temperatures without knowing the details of the magnetic entropy at points between.
Thus, despite the sharp transition in this case, a larger number of magnetic isotherms (including cooling loops below the transformation) was not required for determining the RCPs.

We now consider more generally the RCP obtainable in similar alloy systems with the transformation temperature $T_m$ closer to $T_C$.
As we established, the austenite phases in our Ni-Mn-In materials can be described very accurately as Curie-law paramagnetic with each Mn having a spin fitted to $J=2.0$, and ferromagnetic correlations corresponding $T_C$'s listed in Table~\ref{comptable}.
Except for the critical region very close to $T_C$ in small fields, this gives a good approximation for the spin polarization of samples A and B. For sample C with the structural transformation falling just below the magnetic $T_C$ of the austenite, we extended this analysis into the ferromagnetic-ordered phase by using a mean-field description of the magnetization.

\begin{figure}
\includegraphics[width=\columnwidth]{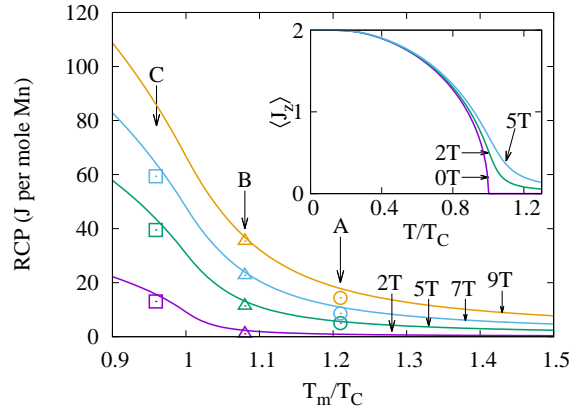}
\caption{Relative cooling power (in unit of J per mole Mn) vs. martensitic transformation temperature calculated for Ni-Mn-In compositions with different $\sfrac{T_m}{T_C}$, as described in text.
		Inset: computed austenite average spin moment for indicated fields.}
\label{rcpplot}
\end{figure}

The inset of Fig.~\ref{rcpplot} shows representative spin polarization results obtained following a standard derivation with the mean spin following a Brillouin function containing the applied magnetic field enhanced by the local exchange field\cite{blundell2001magnetism}.
We assumed each Mn to have a local moment with $J=2$ and $g=2$, coinciding with the magnetization fit.
The magnetization is obtained from the average spin using the known moment per Mn, and
no adjustable parameters are needed to obtain $\int MdH$, which is the integral appearing in Eq.~\ref{rcpmh}.
Fig.~\ref{rcpplot} shows RCP results obtained this way vs. $\sfrac{T_m}{T_C}$, where the temperature at which the integration is evaluated is assumed to match $T_m$ for different compositions along the horizontal axis.
In this calculation we also assumed that the martensite-phase contribution to RCP remains unchanged.
In practice this term may vary in different materials, however, this term is commonly small in inverse magnetocaloric materials.
The open symbols on the graph represent the RCP extracted from specific heat and magnetization for sample A, B, and C.

A good quantitative agreement between the experimental results in open symbol and our simulated solid curves are shown in Fig.~\ref{rcpplot}.
The predicted RCP enhancement based on the previous model has achieved by reducing $\sfrac{T_m}{T_C}$ through varying the composition while the maximal magnetic-induced entropy of materials is actually reduced.

\section{Conclusion}
In this study, we have prepared materials with three compositions and two of them show the only paramagnetic austenite phase while the other shows ferromagnetic austenite phase.
As the Mn concentration decreases, the magnetic stability of the austenite phase is enhanced while structural stability of the martensite phase is reduced,
   which leads to the change of the ratio between the martensitic transition temperature and austenite Curie temperature (i.e. $\sfrac{T_m}{T_C}$) from 1.19 to 0.93.
The relative cooling power (RCP) enhancement is achieved by tuning the magneto-structural stability through reducing $\sfrac{T_m}{T_C}$ and we confirmed the quantitative validity of the analytic RCP evaluation as a function of $\sfrac{T_m}{T_C}$ reported previously\cite{Chen2016176}.
Note that the reduction of $\sfrac{T_m}{T_C}$ ratio resulted in the enhancement of the relative cooling power but it leads to the reduction of the maximal magnetic induced entropy in the same time.
Therefore, compromising consideration should be made during the design of working material for practical applications.

\section*{Acknowledgement}
This material is based upon work supported by the National Science Foundation under Grant No. DMR-1108396, and by the Robert A. Welch Foundation (Grant No. A-1526).
This research used resources of the Advanced Photon Source, a U.S. Department of Energy (DOE) Office of Science User Facility operated for the DOE Office of Science by Argonne National Laboratory under Contract No. DE-AC02-06CH11357.


\bibliographystyle{elsarticle-num}
\bibliography{elsarticle}

\end{document}